\documentclass[aps,twocolumn,pra,superscriptaddress,longbibliography]{revtex4-2}
\usepackage{amsmath}
\usepackage[T1]{fontenc}
\usepackage{graphicx}
\usepackage{amssymb}
\usepackage{amsfonts}
\usepackage{color}
\usepackage{xfrac}

\usepackage{mathtools}

\def\beq{\begin{equation}}
\def\eeq{\end{equation}}

\begin{document}

\title{Energy-efficient neural network inference with microcavity exciton-polaritons}
\author{M. Matuszewski}
\email{mmatu@ifpan.edu.pl}
\affiliation{Institute of Physics, Polish Academy of Sciences, Al. Lotnik\'ow 32/46,PL-02-668 Warsaw, Poland}
\author{A. Opala}
\affiliation{Institute of Physics, Polish Academy of Sciences, Al. Lotnik\'ow 32/46,PL-02-668 Warsaw, Poland}
\author{R.~Mirek}
\affiliation{Institute of Experimental Physics, Faculty of Physics, University of Warsaw, ul. Pasteura 5, PL-02-093 Warsaw, Poland}
\author{M.~Furman}
\affiliation{Institute of Experimental Physics, Faculty of Physics, University of Warsaw, ul. Pasteura 5, PL-02-093 Warsaw, Poland}
\author{M.~Kr\'ol}
\affiliation{Institute of Experimental Physics, Faculty of Physics, University of Warsaw, ul. Pasteura 5, PL-02-093 Warsaw, Poland}
\author{K.~Tyszka}
\affiliation{Institute of Experimental Physics, Faculty of Physics, University of Warsaw, ul. Pasteura 5, PL-02-093 Warsaw, Poland}
\author{T. C. H. Liew}
\affiliation{Division of Physics and Applied Physics, Nanyang Technological University 637371, Singapore}
\author{D.~Ballarini}
\affiliation{CNR NANOTEC-Institute of Nanotechnology, Via Monteroni, 73100 Lecce, Italy}
\author{D.~Sanvitto}
\affiliation{CNR NANOTEC-Institute of Nanotechnology, Via Monteroni, 73100 Lecce, Italy}
\affiliation{INFN, Sez. Lecce, 73100 Lecce, Italy}
\author{J.~Szczytko}
\affiliation{Institute of Experimental Physics, Faculty of Physics, University of Warsaw, ul. Pasteura 5, PL-02-093 Warsaw, Poland}
\author{B.~Pi\k{e}tka}
\affiliation{Institute of Experimental Physics, Faculty of Physics, University of Warsaw, ul. Pasteura 5, PL-02-093 Warsaw, Poland}

\begin{abstract}
 We propose  all-optical neural networks characterized by very high energy efficiency and performance density of inference. We argue that the use of microcavity exciton-polaritons allows to take advantage of the properties of both photons and electrons in a seamless manner. This results in strong optical nonlinearity without the use of optoelectronic conversion. We propose a design of a realistic neural network and estimate energy cost to be at the level of attojoules per bit, also when including the optoelectronic conversion at the input and output of the network, several orders of magnitude below state-of-the-art hardware implementations. We propose two kinds of nonlinear binarized nodes based either on optical phase shifts and interferometry or on polariton spin rotations. 
\end{abstract}

\maketitle

\section{Introduction}

The progress in communications, information processing, and mobile technologies resulted in the advent of the era of big data. This received an incredible burst with the developments in parallel computing hardware, and in particular in artificial intelligence and neural networks (NNs)~\cite{LeCun_DeepLearning,Misra_ANNSurvey}. Practical applications of machine learning quickly became an important part of the economy, and currently include, among others: natural language processing, image and sound recognition, autonomous vehicles, finance, marketing, and research. At the same time, these developments put a strain on computing systems, which have to process data faster and more efficiently than ever before. This becomes a serious issue as the energy consumption of information processing and communication systems is set to surge. It is already a significant part of the global energy consumption, and is expected to reach over 20\% of global electricity use by 2030~\cite{Andrae_Electricity}, becoming one of the main bottlenecks of further progress. This problem has been recognized by the machine learning community~\cite{Strubell_Ganesh_McCallum_2020,Canziani,Li_SustainableComputing}.

Meanwhile, the increase of performance of complementary metal-oxide semiconductors (CMOS) saturates, significantly deviating from Moore's law~\cite{Waldrop_Moore}. In recent years, the shift to architectures with many  parallel computing units, such as GPUs or TPUs, has been the dominant trend, but this avenue is limited by Amdahl's law~\cite{Kitayama_PhotonicAccelrator}. In result, much effort has been dedicated to research on possible alternatives to the CMOS technology for information processing~\cite{Grollier_review}. In particular, there have been great advancements in machine learning with photons, in both all-optical and optoelectronic systems~\cite{bogaerts2020programmable,Wetzstein_review,Shastri_review}. Photonic information processing benefits from high speed, parallelization, low communication losses, and high bandwidth. Fully functional photonic neurons including nonlinear activation functions were demonstrated~\cite{tait2019silicon,amin2019ito,Shastri_review,Wetzstein_review}, including spiking neurons~\cite{Tait_LightTech_2014,brunstein2012excitability}, as well as neural networks~\cite{goodman1978fully,Psaltis_Holography,farhat1985optical,lu1989two,Psaltis_Associative,Psaltis_Adaptive,tait2017neuromorphic,hughes2019wave,Feldmann_AllOpticalSpikingNetwork,shi2019deep,Zuo_AllOpticalNN,Vandoorne}. Certain networks achieved high performance in challenging machine learning tasks, such as image and video recognition~\cite{Brunner_ReinforcementLearning,Ballarini_Neuromorphic,Mirek_Neuromorphic,Lin,antonik2019large,antonik2019human,Zhou_LargeScale}. Scalable vector-matrix multiplication operations and convolutions, which are at the core of neural network implementations, were demonstrated using photons~\cite{Soljacic_DeepLearning,Feldmann_parallel,Chang_CNN,Xu_TOPS}. All these remarkable developments indicate that photonic information processing is maturing, and may become a serious competitor for electronics in the near future, as recognized both in the academia and the industry~\cite{Pile_Rise,rodrigues2021weighing}.

The most serious issue taming progress in electronic systems is related to energy consumption and heat generation, which  results in the phenomenon of ``dark silicon''~\cite{Waldrop_Moore,esmaeilzadeh2011dark}. The main source of losses at high data rates is not the cost of the actual logic operations, but communication using electronic channels~\cite{Miller_Attojoule}. On the other hand, communication using photons can be almost lossless. For this reason, electric wiring has been successively replaced by photonic waveguides at long and medium distances, and are now being commercially implemented even at the chip scale~\cite{Kachris_OpticalInterconnects}. However, electronics is still necessary for signal amplification and processing. The drawback of photonics is the weakness of nonlinearity, or effective interactions between photons. Nonlinearity is crucial for the implementation of nontrivial information processing,  either in the form of a transistor or a neuron activation function. All-optical information processing has been unpractical due to high intensity of optical beams required, which effectively results in high energy cost per operation.

Alternatively, since nonlinearity occurs at very low energy levels in electronics, optoelectronic conversion can be used, so that nonlinear transformations are implemented electronically, and photons are used for communication or linear operations only. However, there are still serious issues with the implementation of scalable optoelectronic information processing. Integration of light sources is difficult,  and the spatial scales for efficient electronics (nanometers) and photonics (micrometers) are incompatible. The energy cost of optoelectronic conversion in logical elements is typically such that it overcomes any gain from the use of optics for communication. It is estimated that net energy efficiency benefits can only be expected if the spatial extent of photonic modes is reduced to the nanometer scale~\cite{Miller_Attojoule}. This may be difficult in practice without incurring additional losses, either due to absorption or imperfect confinement of photons at the sub-wavelength scale. Since practical solutions to these issues are yet to be demonstrated, it is desirable to find an all-optical alternative for energy-efficient information processing. Such an approach requires strong optical nonlinearity without optoelectronic conversion. This would allow for fully exploiting the intrinsic ultrashort time scales and high energy efficiency of photonics in complex information processing tasks that require nonlinearity.

We recently demonstrated hardware neuromorphic systems where strong nonlinearity resulted solely from interactions of exciton-polaritons, quantum superpositions of light and matter~\cite{Opala_NeuromorphicComputing,Mirek_Neuromorphic,Ballarini_Neuromorphic}.
Such superpositions, in the form of mixed quasiparticles of photons and excitons~\cite{Kavokin_Book,Carusotto_QuantumFluids}, are characterized by excellent photon-mediated transport properties and strong exciton-mediated interactions~\cite{Walker_UltraLowPowerSolitons,Baumberg_SubfemtojouleSwitches}. Their unique properties led to the discovery of remarkable phenomena, including nonequilibrium Bose-Einstein condensation~\cite{Kasprzak_BEC,Carusotto_QuantumFluids}, superfluid-like states~\cite{Amo_Superfluidity,Marzena_Superfluidity}, interactions of quantum defects~\cite{Sanvitto_TopologicalOrder,Deveaud_VortexDynamics,Sanvitto_InteractionsScatteringVortices}, lasing of topological edge states~\cite{Amo_LasingTopological,Hofling_TI}, as well as demonstration of polariton transistors and switches~\cite{Bramati_SpinSwitches,Sanvitto_TwoFluid,Baumberg_SubfemtojouleSwitches,Sanvitto_Transistor,Lagoudakis_RTOrganicTransistor,Savvidis_TransistorSwitch}, gates~\cite{Baranikov_AllOptical,Mirek_Neuromorphic}, neurons~\cite{Liew_Neurons,Espinosa_perceptrons}, simulators~\cite{Lagoudakis_XYSimulator} and nonlinear phenomena at the femtojoule level~\cite{Walker_UltraLowPowerSolitons,Baumberg_SubfemtojouleSwitches}.
In our proof-of-principle neural network implementations, both photonic and electronic layers were used, but the latter performed linear operations only. These experiments, based on reservoir computing~\cite{Ballarini_Neuromorphic} and binarized neural networks~\cite{Mirek_Neuromorphic}, achieved accuracy of 93\% and 96\% in the MNIST handwritten digit benchmark, respectively. Moreover, the strong polariton nonlinearity, resulting from Bose-Einstein condensation, allowed to achieve energy efficiency of 16 picojoule per synaptic operation (SOP) in an all-optical binarized neuron~\cite{Mirek_Neuromorphic}.

Here, we argue that semiconductor microcavity systems can be used to construct fully functional, all-optical neural networks characterized by extremely high energy efficiency of inference. We show why using polaritonics in place of standard nonlinear optical phenomena, is the key to achieving such a performance. 
Exciton-polaritons can perform information processing that takes advantage of the properties of both photons and electrons in a seamless manner. We propose a design of a polariton-based neural network in which fast, parallel, and low-loss communication using photons is accompanied by the strong nonlinearity induced by exciton interactions, without the need for optoelectronic conversion. We estimate the performance of such a network in the case of resonant excitation, and predict that nonlinear inference at attojoules per bit energy cost can be achieved, outperforming current technologies by orders of magnitude. We propose how a simple network, based on nonlinear binarized nodes, could be implemented, taking advantage of either the interference of photons or polariton spin rotations. We estimate two key measures of the proposed design -- the energy efficiency and performance density.
Our estimations indicate that using currently available optical elements, the network could reach the energy efficiency of 4$\times$10$^{16}$ SOP\,s$^{-1}$W$^{-1}$ (synaptic operations per second per watt) and performance density of 10$^{16}$ SOP\,s$^{-1}$ mm$^{-2}$ (synaptic operations per second per millimeter squared), which are higher than the limits of the current semiconductor technology~\cite{Xu_EdgeInference}, as well as other technologies under investigation~\cite{ITRS_BeyondCMOS}, by four and three orders of magnitude, respectively.  Finally, we discuss the challenges and limitations of the proposed platform, including those related to optical losses, and indicate the possible applications where polaritonics may compete with CMOS electronics.

\section{General estimates} \label{sec:estimates}

\subsection{Exciton-polaritons}

\begin{figure}
    \includegraphics[width=0.95\linewidth]{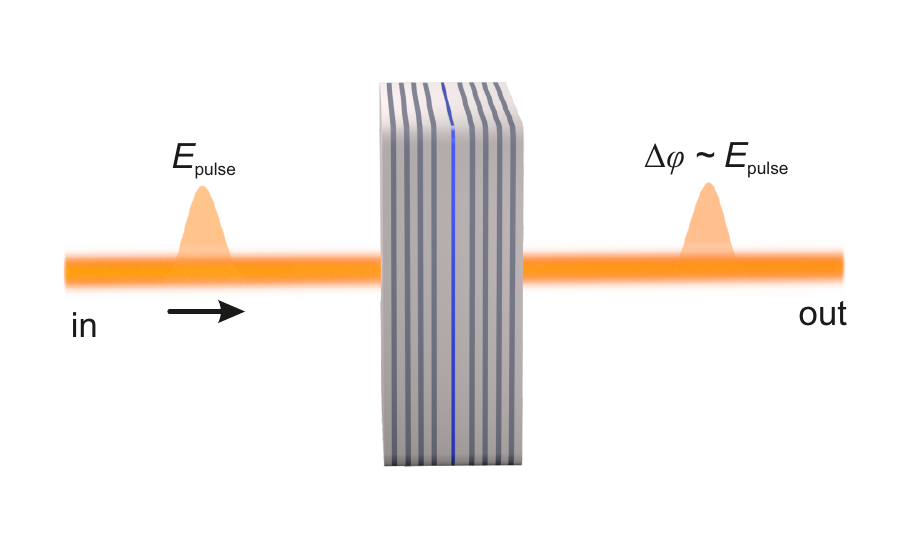}
    \caption{Simple setup for the consideration of energy efficiency. The microcavity contains exciton-polariton quantum wells or a nonlinear optical medium. The optical phase shift of the output pulse depends on the energy of the input pulse. This nonlinear (i.e.~optical intensity dependent) transformation is useful for information processing when phase shifts of the order of $\pi$ can be achieved.}
    \label{fig:fig1}
\end{figure}

To estimate the theoretical minimum of energy required to perform a single nonlinear operation using exciton-polaritons, we consider the simple device depicted in Fig.~\ref{fig:fig1}. The optical phase of the output pulse depends on the input pulse intensity due to nonlinear dynamics of polaritons inside the microcavity. We assume that the nonlinear process is third-order, i.e.~the phase shift depends linearly on the intensity, and results from the repulsive interactions between polaritons mediated by their exciton component~\cite{Carusotto_QuantumFluids}. While other processes, including interactions with the reservoir of weakly coupled excitons, or the reduction of light-matter coupling strength, can also lead to a similar nonlinear response, we consider direct polariton-polariton interactions as the process which is likely to play the most important role at the shortest timescales. 

We assume that to implement a useful nonlinear transformation, the phase shift induced by the nonlinearity should be of the order of $\pi$. Such a phase shift can be, for example, subsequently converted into full-scale amplitude modulation using a Mach-Zender interferometer. Therefore, it is enough to construct optical components that turn on or off transmission through the device depending on the nonlinear phase shift. However, we do not exclude the possibility that smaller phase shifts can be also used for computation.  We compare the natural energy scales to estimate the number of photons necessary to realize such a phase shift. This leads to the formula $n g  > \hbar \gamma$ where $n$ is the quantum well exciton density, $g$ is the exciton-exciton interaction coefficient, and $\gamma$ is the polariton decay rate, which is the inverse of lifetime of polaritons in the cavity. Here  $n \approx \chi N_{\rm photons} /(N_{\rm QW} S)$, where $N_{\rm photons}$ is the number of photons which were converted into polaritons, $\chi$ is the exciton Hopfield coefficient, $S$ is the area of the illuminated surface, and $N_{\rm QW}$ is the number of quantum wells accommodating excitons. To estimate a realistic lower limit of energy consumption, we consider resonant excitation configuration, in which polaritons are created directly by a picosecond laser pulse. Such a configuration is much more energy efficient than the nonresonant excitation used in the previous experiment~\cite{Mirek_Neuromorphic}, since most of the photons can be converted directly into polaritons. As we are concerned with order of magnitude estimations, for the time being we neglect the photons reflected or absorbed by the microcavity or other optical components. More precise estimates will be presented in Sec.~\ref{sec:total}.

In the theoretical estimation of energy efficiency we use realistic microcavity parameters. Following references~\cite{Estrecho_DirectMeasurementInteractionStrength,Snoke_BECLongLifetime} we use the values of parameters $g=2\,\mu$eV$\mu$m$^2$, $\gamma=(270\,$ps$)^{-1}$. These parameters correspond to state-of-the-art GaAs microcavities, but a range of inorganic materials including CdTe, ZnO, GaN, and perovskites are expected~\cite{Tassone_ExcitonScattering} to be characterized by exciton interaction coefficients $g$ in the range $2-5\,\mu$eV$\mu$m$^2$. Many of them were used to observe exciton-polaritons at room temperature~\cite{Grandjean_RTPolaritonLasing,Fieramosca_perovskites,Fieramosca_perovskites_new,Su_Perovskites,Malpuech_ZnOCondensate}. In this work we assume a room temperature operation to avoid the energy cost of cooling. We assume that a single quantum well is present in the microcavity, and use a conservative estimation of the surface area of a single polariton gate $S=10\,\mu$m$^2$. Note that polariton modes as small as 1\,$\mu$m$^2$ have been realized experimentally~\cite{Volz_QuantumCorrelatedPhotons}, but structuring light on such a small length scale could be challenging due to interference effects. 

This gives the estimate of the minimum energy of a single input pulse $E_{\rm pulse}=N_{\rm photons}E_{\rm photon}$ at the level of 3 attojoules ($3\times 10^{-18}$\,J), which corresponds roughly to  12 photons per pulse, for the photon energy $E_{\rm photon}=1.6$\,eV. One may worry that too small a photon number introduces errors due to quantum fluctuations; these will be discussed in Sec.~\ref{sec:quantum}. Since a simple binary neuron (see Sec.~\ref{sec:methods}) requires 4 pulses and performs 2 synaptic operations, the energy efficiency bound is estimated as $1.7\times10^{17}$ SOP\,s$^{-1}$W$^{-1}$. This can be compared, for example,  with the energy efficiency of $4\times 10^{10}$\,SOP\,s$^{-1}$W$^{-1}$ obtained with the IBM TrueNorth neuromorphic chip~\cite{Merolla}.

\subsection{Nonlinear optical media}

Microcavities can be used to construct nonlinear elements without the use of exciton-polaritons. Instead, any medium exhibiting third-order nonlinearity can be embedded in a microcavity to realize the same concept. In this case, the energy required for a single operation can be estimated from the strength of optical phase modulation, knowing the parameters $n_2$ (nonlinear Kerr index), $n_0$ (refractive index) and $n_g$ (group refractive index). The third-order nonlinearity results in the intensity-dependent refractive index change $\delta n = n_2 I$ = $n_2 P/S$, where $P$ is the power and $S$ is the surface area. The resulting change of resonator frequency is $\delta \omega =- (\delta n / n) \omega$ as the wavelength is fixed by the cavity size and the number of antinodes. The corresponding shift of resonator energy is $\delta E = \hbar \delta \omega$. The minimum energy shift required for the operation of a gate can be estimated as in the case of polaritons, by comparing it to the natural energy scale of the cavity $|\delta E| = \hbar \gamma$. If this condition is fulfilled, photons can acquire a $\pi$ phase when interacting in the microcavity.

Finally, the energy of the input pulse can be estimated from $E_{\rm pulse}=P  t_{\rm cav}$, where $t_{\rm cav}$ is the time necessary for the light to perform a round-trip inside the cavity, $t_{\rm cav}=2Ln_g/c$, where $L$ is the length of the cavity. This leads to the formula $E_{\rm pulse}=2 n_0 n_g \gamma S L /(n_2 \omega c)$ for the energy of the input pulse.
We estimate that materials exhibiting strong and fast nonlinearity, such as silicon or gallium arsenide semiconductors in the weak coupling regime, with $n_2$ of the order of $10^{-17}$ m$^{2}$ W$^{-1}$, would require at least $10^{-14}$ J (10 fJ) per synaptic operation, four orders of magnitude higher than in the case of exciton-polaritons. At the same time, materials which exhibit slow nonlinear processes, such as photo-refractive media, would not be energy efficient due to long illumination times required to build up the nonlinear index change.

\subsection{Optoelectronic approach}

Electronic devices are excellent for nonlinear operations, with energy per bit in a logical gate reaching possibly the level of 50-100 aJ~\cite{Miller_Attojoule,ITRS_Executive}. However, communication on chip requires charging electrical wires to the level of  1 V, because lower voltage leads to strong leakage current in transistors. The energy cost of charging communication lines results in large energy dissipation, of the order of pJ per bit for floating point operations, memory access, and even larger cost for off-chip communication.

On the other hand, optoelectronic approach, where nonlinear operations are performed electronically, but communication is realized with light, is a possible solution that takes advantages of both methods. However, this requires optoelectronic conversion for each information bit, which itself generates pJ energy cost in typical devices~\cite{Kachris_OpticalInterconnects}. Since the cost of optoelectronic conversion scales with size, it was suggested that net benefits can be expected if light modes are confined to volumes at the nanometer scale~\cite{Miller_Attojoule}. This may be difficult in practice, as the confinement of light modes to below-wavelength spatial scales is usually associated with strong losses either due to absorption by metallic mirrors, or imperfect confinement of dielectric mirrors.

\section{All-optical neural network}

In this section we discuss a possible implementation of polariton nonlinearity in an all-optical neural network. A single hidden layer network of binarized nodes is considered for its simplicity. Here and in the following, we use the term ``binarized'' to describe a neural network with binary inputs and activations in the hidden layer. While numerous optical neural networks or machine learning systems have been described in the literature, only~\cite{Ballarini_Neuromorphic,Mirek_Neuromorphic} considered the use of exciton-polaritons.

\subsection{Design}

\begin{figure*}
    \centering
    \includegraphics[width=0.70\linewidth]{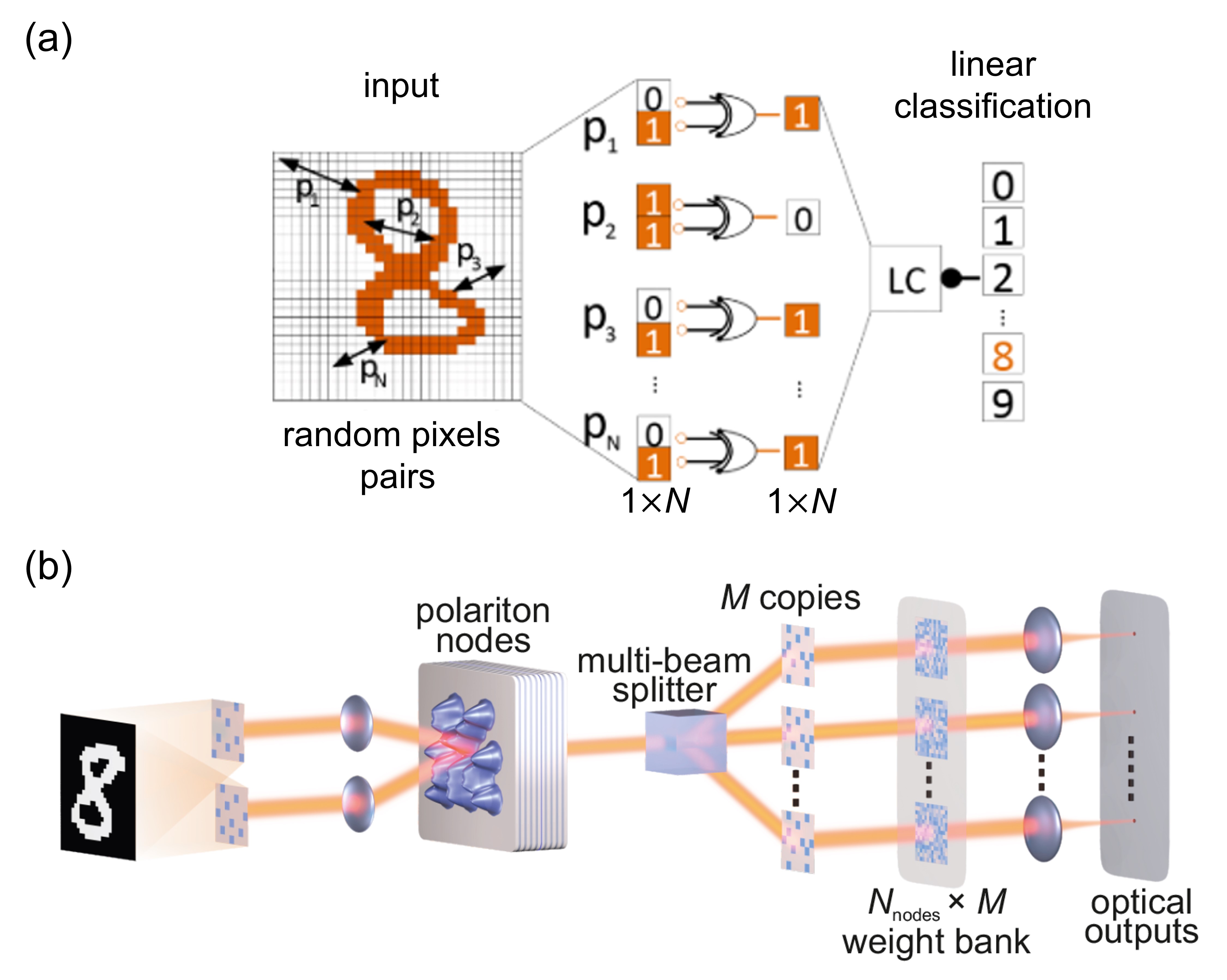}
    \caption{(a) Conceptual scheme of a single hidden layer network and (b) a possible all-optical implementation. The raw input (schematically depicted as a handwritten digit) is spatially encoded with light pulses which are assumed to arrive simultaneously. It is split into two parts visualized with the dotted arrays, which encode the two inputs of multiple XOR gates. Each of the gates corresponds to a single pixel of the array. The transmitted intensity contains the result of the XOR gates if one of the methods of Sec.~\ref{sec:methods} is used. This is subsequently split into $M$ copies which correspond to the number of classes. The weight banks implement linear classification of the result in the $N_{\rm nodes}$-dimensional feature space. Light intensity at the output measured by the detectors contains information about the probability that the sample belongs to a certain class.}
    \label{fig:design}
\end{figure*}

We propose a design of a simple all-optical neural network taking advantage of the polariton nonlinearity. There are no electrical or optoelectronic elements in the system and both inputs and outputs are assumed to take the form of optical pulses. This makes the proposal distinct from previous experimental implementations~\cite{Mirek_Neuromorphic,Ballarini_Neuromorphic}, and is crucial for the energy efficiency of the network. Moreover, all elements of the network  such as beam splitters, mirrors, lenses, light filters are passive and do not require an external power supply. In contrast to experiments realized in exciton-polariton systems~\cite{Ballarini_Neuromorphic,Mirek_Neuromorphic}, our design does not involve optoelectronic conversion or electronic information processing at any element within the network. Although we consider inputs and outputs as optical, we appreciate that they may have to be converted from or to electronic signals for compatibility with other systems. We account for the corresponding additional energy consumption in Sec.~\ref{sec:total}. The proposed network is one of the simplest possible applications of polaritons for optical information processing, but not the only one possible, and many other configurations can be envisaged. We discuss some of the possible extensions in Sec.~\ref{sec:other}. The aim of this work is to propose a simple design to indicate the potential of polaritonics for energy efficient computing.

We consider a large network of all-optical binarized neurons where inputs and activations in the hidden layer are two-level. Such a network can be used for complex classification tasks~\cite{Mirek_Neuromorphic,Bengio_Binarized,Rastegari}. The logical structure of the network is analogous to the one introduced in Ref.~\cite{Mirek_Neuromorphic} and shown schematically in Fig.~\ref{fig:design}. In~\cite{Mirek_Neuromorphic}, we experimentally realized a single binarized neuron using a polariton microcavity excited with a nonresonant laser pulse, achieving 16 pJ per bit energy efficiency. Here, we propose to use resonant excitation to achieve the same functionality, but at much higher energy efficiency, since resonant pumping selectively excites polaritons and allows most of the input photons to be converted to polaritons. The activation function of neurons is such that they effectively perform XOR logical gate operations. In the case of information encoding with light intensity, the intensity levels corresponding to ``0'' and ``1'' bits can be defined arbitrarily and differently at the input and output of the gates, as long as they are consistent between the gates. We consider a single hidden-layer network, composed of $N_{\rm nodes}$ polaritonic gates. The linear classification at the output layer is realized by an all-optical vector-matrix multiplier. Optical vector-matrix multiplication has been realized in many experiments~\cite{Brunner_ReinforcementLearning, Zuo_AllOpticalNN, Chang_CNN, farhat1985optical, goodman1978fully,gruber2000planar, lu1989two}, and recently multiplication of 3000 variables has been demonstrated~\cite{Spall_OVMM}.

The input information is encoded in space, and all input pulses are assumed to arrive at the same time. In the particular example considered, pixels from the MNIST dataset are binarized, and assigned to elements of two equally sized arrays. We assign random pairs of pixels to the elements of the two arrays, as depicted in Fig.~\ref{fig:design}. The same pairs of pixel positions denoted by $p_1\ldots p_n$ are assigned to the same array elements $1\ldots n$ for all digits. This allows the network to detect nontrivial correlations between distant pixels even in the single-layer XOR network. The above stage does not require any nonlinear operation and can be implemented all-optically, for example using a diffractive optical element (DOE) or a three-dimensional laser-written waveguide array. The arrays of pulses, forming the two  binary inputs of the gates, are directed at the same area of the sample, creating an array of $N_{\rm nodes}$ nonlinear nodes. The optical output (light transmission) from each of the nodes corresponds to the result of an XOR operation. To obtain an XOR activation function of the nodes, we propose to use one of the methods described in Sec.~\ref{sec:methods}. Further on, a vector-matrix multiplication is realized with linear optical elements. Precisely, each of the $N_{\rm nodes}$ pulses which constitute output of the nodes is split into $M$ copies (for example using a diffractive optical element), where $M$ is the number of classes. These copies are directed at an $N_{\rm nodes} \times M$ optical filter array, which applies synaptic weights by selectively attenuating intensities of the pulses. The resulting weighted pulses are combined into $M$ outputs.
The output of the device consists of $M$ pulses which are the result of the multiplication. The optical vector-matrix multiplication has been realized both in the case of coherent and incoherent light~\cite{Brunner_ReinforcementLearning, Zuo_AllOpticalNN, Chang_CNN, farhat1985optical, goodman1978fully,gruber2000planar, lu1989two,Spall_OVMM}. The intensity at the output is related to the probability that the input sample belongs to the particular class. This intensity can be measured by a detector such as a photodiode to convert it to an electrical signal.

The optical elements necessary to build the above system are readily available. The input can be created using a laser generating picosecond optical pulses, and input information can be encoded with arrays of micrometer-sized optoelectronic modulators~\cite{Nozaki_Femtofarad,Watts_Ultralow,sorger2017scaling}. The multi-beam-splitter used for the vector-matrix multiplication can be either implemented with an appropriately shaped diffractive optical element or a system of standard beam splitters. Finally, arrays of sensitive photodetectors can be used to detect the output signals~\cite{Nozaki_Femtofarad}. The optical weight banks can be implemented with any form of a tunable semi-transparent filter, for example using programmable nanophotonics~\cite{Soljacic_DeepLearning,bogaerts2020programmable}, a liquid crystal array, or a phase change material array~\cite{Feldmann_AllOpticalSpikingNetwork}.

\subsection{Numerical results}

We simulate the proposed system, assuming that $N_{\rm nodes}=8 000$ gates are characterized by the same level of inaccuracy in terms of shot-to-shot variance of intensity as in the experiment~\cite{Mirek_Neuromorphic}. Despite the relatively large amount of experimental noise in that experiment, we find that such a network is able to recognize MNIST handwritten digits with 95.9\% accuracy, only 0.1\% lower than the network of ideal XOR gates. This is close to the 96\% accuracy on the MNIST task of the state-of-the-art for hardware neuromorphic implementations~\cite{Qian_MemristorCNN,Chen_ClassificationSilicon,Mirek_Neuromorphic} , although sophisticated software simulations of neural networks can reach much higher accuracy of recognition, at the cost of much longer operation time and higher energy consumption~\cite{LeCun_MNIST}.

\subsection{Total energy consumption}\label{sec:total}

The estimations of energy efficiency presented in Sec.~\ref{sec:estimates} have to be treated as theoretical lower bounds for the energy efficiency of an ideal system. Here, we attempt to take into account all sources of energy loss and estimate more realistic energy efficiency including all necessary optical and optoelectronic components. Until now we have only taken into account the optical energy of input pulses. However, most of the information in the modern world is carried electronically, with the exception of optical fibers connections. In the case when information is encoded electronically, we expect that the conversion of electronic to optical signal can lead to a significant overhead in terms of energy efficiency and speed. In particular, commercial ultrashort pulse lasers reach wall-plug efficiencies, or conversion ratio of electric to optical energy, above 10\%.

We present a calculation of the total energy consumption of a complete network, taking into account losses occurring at all elements of the system, i.e.~laser source, modulators, microcavity, weight banks, and photodetectors. We consider two cases: (A) an ``idealized'' large scale system, with parameters corresponding to state-of-the-art optical elements, and (B) a proof-of-principle system with a relatively small number of nodes and accessible optical elements.

The average energy per synaptic operation in a binarized network can be calculated as $E_{\rm op} = E_{\rm total} / (2 N_{\rm nodes})$, where $E_{\rm total}$ is the total energy consumed by the system and $N_{\rm nodes}$ is the number of binary nodes (gates) in the hidden layer, each gate with two synaptic inputs. The total energy consumption for a complete neural network is
\beq
E_{\rm total}= E_{\rm source} + E_{\rm modulators} + E_{\rm detectors},
\eeq
where $E_{\rm modulators}$ is the net electrical energy required to supply optoelectronic modulators that encode input data, and $E_{\rm detectors}$ is the energy required to supply photodetectors at the output layer. In the proposed architecture, all data is injected, processed, and read out in parallel. The number of modulators is equal to the number of input bits, and the number of photodetectors is equal to the number of classes.
The microcavity and the weight bank are passive elements, and do not appear directly in the formula above, but the optical energy loss occurring in these elements will influence the required laser power.

The energy of the source laser will be bounded from below by two main factors. It has to be high enough for the operation of all the polariton nodes in the hidden layer, and on the other hand, provide sufficiently high optical energy for the photodetectors in the output layer. Of the two constraints, the first one will be typically the more strict one, due to the large number of nodes and much lower number of classes measured by photodetectors in a standard network. We estimate the electrical energy required to supply the laser as
\beq
E_{\rm source} \geq \frac{E_{\rm nodes}}{\eta_{\rm L}} = \frac{2N_{\rm nodes}E_{\rm pulse}}{\eta_{\rm C}\eta_{\rm L}}
\eeq
where $E_{\rm nodes}$ is the optical energy required to supply all the polariton nodes, $\eta_{\rm L}<1$ is the wall-plug efficiency of the laser, $E_{\rm pulse}$ is the minimum energy per pulse as estimated in Sec.~\ref{sec:estimates}, $\eta_{\rm C}<1$ is the microcavity transmission coefficient (ratio of output/input energy transmitted for ``1'' output logic value). We overestimate the energy by assuming that the energy of polaritons interacting in the microcavity is equal to the energy of pulses transmitted through the microcavity in the ``1'' output state. We neglect losses at optical elements such as beam splitters and mirrors, as they will contribute marginally to the final efficiency. We can estimate the total energy per operation as
\beq
E_{\rm op}\geq \frac{E_{\rm pulse}}{\eta_{\rm C}\eta_{\rm L}} + \frac{N_{\rm mod}}{N_{\rm nodes}}E_{\rm mod} + \frac{N_{\rm det}}{N_{\rm nodes}}E_{\rm det},
\eeq
where $N_{\rm mod}$ is the number of modulators, $E_{\rm mod}$ is the energy cost per bit for a single modulator,  $N_{\rm det}$ is the number of detectors, and $E_{\rm det}$ the energy cost per bit for a single detector. 
The above formula describes the important fact that if a single bit of information is shared as input by many nodes in the hidden layer, which is a typical situation in neural networks, the cost of the input optoelectronic conversion will be divided between all the nodes that use this bit of information. This can reduce the energy cost of generating an input pulse for a particular hidden node by orders of magnitude with respect to the energy cost of the optoelectronic switch. Similarly, the energy of detectors per operation will be proportionally reduced.

The conversion of information from electronic to optical signal can be realized with ultra-efficient modulators, which reach energy efficiency of several femtojoules per bit~\cite{Nozaki_Femtofarad,Watts_Ultralow}. Polariton spin switches were also recently demonstrated to achieve energy efficiency at the femtojoule level~\cite{Baumberg_SubfemtojouleSwitches}. The efficiency of photodetectors has also reached very high (femtojoule) levels~\cite{Nozaki_Femtofarad}. The number of photodetectors $N_{\rm det}$ may be very low. For example, if the task of the network is to distinguish between two classes of objects, only two photodetectors will be necessary, and consequently their share in the total energy budget will be negligible. 

A single laser pulse can be efficiently redistributed among inputs. It can be split with DOEs or beam splitters into multiple copies, and used to generate multiple bits of information. A single pulse can even be split and redirected into many devices if its energy is high enough to power many of them, so the energy of a single laser pulse should not be treated as the bound for efficiency of a network. Likewise, if the number of classes is large, sensor arrays incorporating many elements may be used to detect thousands of optical outputs simultaneously, which can make the optoelectronic conversion more efficient.

\begin{table}
\begin{tabular*}{\columnwidth}{@{\extracolsep{\fill} } l c c } 
 \hline
  & case A & case B \\
 \hline \hline
 $N_{\rm inputs}$ & 10 000 & 10 \\ 
 $N_{\rm outputs}$ & 10 000 & 2 \\ 
 $N_{\rm nodes}$ & $10^6$ & 100 \\ 
 Surface area of a single node & $1\,\mu$m$^2$ & $10\,\mu$m$^2$ \\ 
 Polariton lifetime $\gamma^{-1}$ & 100 ps & 1 ps \\
 Cavity transmission coefficient $\eta_{\rm C}$ & 0.9 & 0.1 \\
 Laser wall-plug efficiency $\eta_{\rm L}$ & 0.2 & 0.1 \\
 $E_{\rm modulator}$, $E_{\rm detector}$ & 1 fJ & 1 pJ \\
 \hline
 Energy cost per operation $E_{\rm op}$ & 24 aJ & 200 fJ \\
 Energy efficiency (SOP\,s$^{-1}$W$^{-1}$) & $4 \times 10^{16}$ & $5 \times 10^{12}$ \\
 \hline
\end{tabular*}
\caption{Parameters of the two  devices considered in the text and the total energy cost per synaptic operation.}
\label{table}
\end{table}

In Table~\ref{table} we show detailed estimations for two cases, corresponding to an ``idealized'' case A with parameters of state-of-the-art components and microcavities, used to solve a complicated machine learning task (such as ImageNet image recognition) and a ``proof-of-principle'' case B with parameters closer to off-the-shelf optical components and not yet very optimized microcavities in which room temperature polaritons have been already observed~\cite{Grandjean_RTPolaritonLasing,Malpuech_ZnOCondensate,Fieramosca_perovskites}. We consider polaritons with realistic interaction constant~\cite{Tassone_ExcitonScattering} $g=3 \mu$eV$\mu$m$^2$. In the idealized case A, we assume the energy cost of modulators and photodetectors to be of the order of 1 fJ per pulse as in Ref.~\cite{Nozaki_Femtofarad}, cavity polariton lifetime 100 ps and transmission coefficient of 90\% (see Sec.~\ref{sec:supertransparent}) and 20\% wall-plug efficiency of the laser. A large scale neural network with a million nodes, 10 000 inputs and 10 000 output classes is estimated to reach energy efficiency of $E_{\rm op}=24$ aJ per synaptic operation. In the ``proof-of-principle'' case B, we assume the energy cost of modulators and detectors at 1~pJ, 10\%~laser wall-plug efficiency, 10\%~cavity transmission coefficient and 1~ps polariton lifetime. A simple network with 10 inputs, 100 nodes and 2 output classes achieves efficiency of $E_{\rm op}=200$ fJ per operation, which is two orders of magnitude below electronic neuromorphic realizations~\cite{Merolla} and one order of magnitude below the state-of-the-art specialized electronic neural network accelerator~\cite{Chen_DianNao}.

For completness, we check if the output contains enough optical energy for photodetectors. A commercial photodetector characterized by noise equivalent power of 6 ${\rm pW}/\sqrt{\mathrm{Hz}}$, is capable of detecting a signal of a 130 nW optical power within a 200 MHz bandwidth at a signal to noise ratio equal to one. This corresponds to the required optical energy of 650 aJ per sample in each photodetector. Considering the high ratio of the number of optical nodes to the number of detectors, we conclude that the output contains enough optical energy in both case A and case B. However, in the case of tasks that require a high bit depth, such as regression, the photodetector sensitivity may become the most important limiting factor of energy efficiency.

\subsection{Comparison with other systems}

\begin{figure}
    \includegraphics[width=\linewidth]{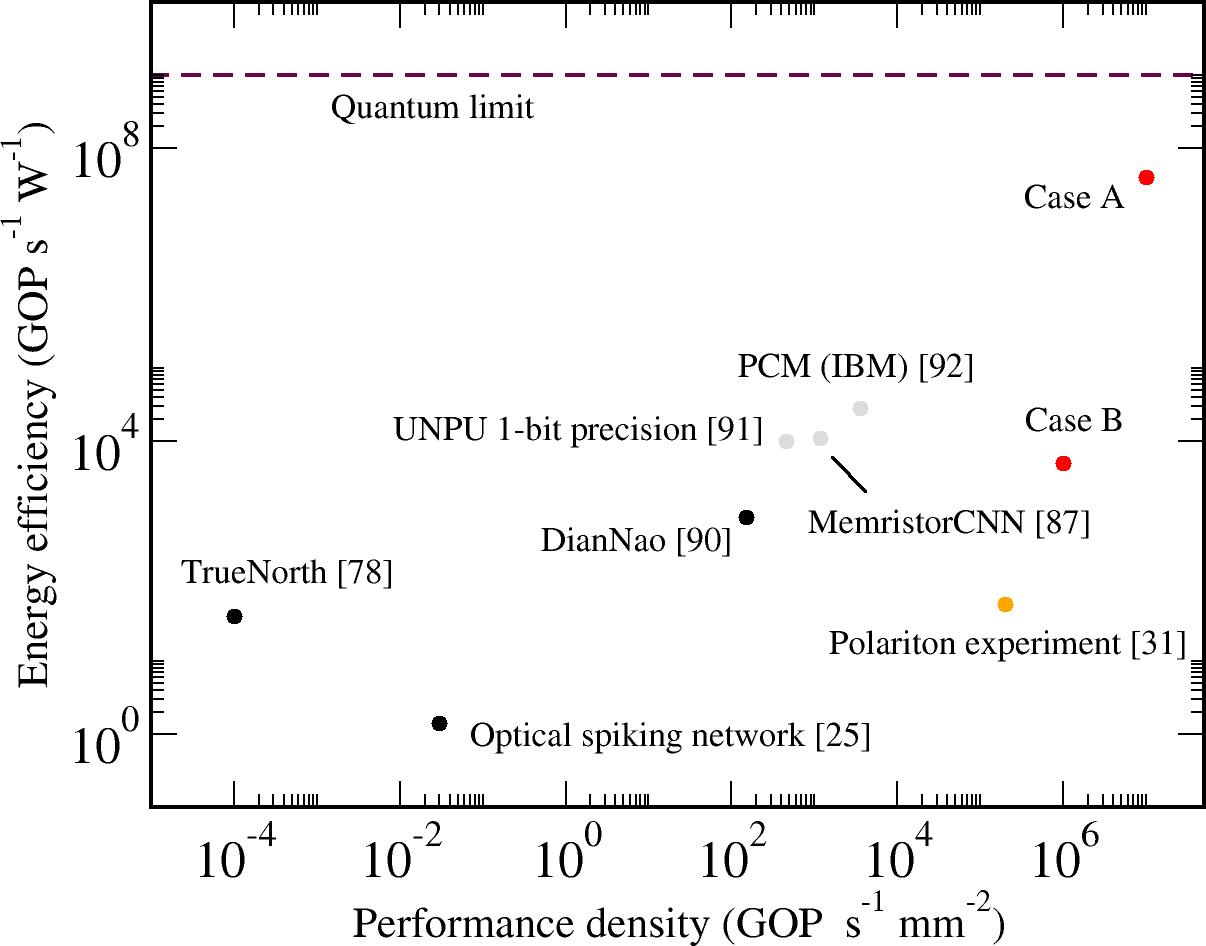}
    \caption{Comparison of energy efficiency and performance density estimates of polariton systems (red dots) and other systems~\cite{Feldmann_AllOpticalSpikingNetwork,Chen_DianNao,Mirek_Neuromorphic,Lee_UNPU,Merolla,Ambrogio_EquivalentAccuracyNN,Qian_MemristorCNN}. Grey points correspond to systems for which estimates of only a certain part of the network was provided~\cite{Lee_UNPU,Qian_MemristorCNN}, or the implementation is partly realized by software~\cite{Qian_MemristorCNN,Ambrogio_EquivalentAccuracyNN}. Orange dot corresponds to the experiment with nonresonantly pumped polaritons~\cite{Mirek_Neuromorphic}. Dashed line is the single-photon quantum limit for optics.}
    \label{fig:comparison}
\end{figure}

We first compare the inference efficiency of our network and non-neuromorphic systems that implement software simulations of neural networks. We consider the most energy-efficient supercomputer from the Green500 list, which performs a single floating point operation (FLOP) operation at the cost of approximately 60 pJ. We assume that in a computer simulation of an artificial neural network a single operation roughly corresponds to a single synaptic event, such as accumulation of a single weighted input, 1 FLOP $\approx$ 1 SOP. This gives the energy efficiency of $1.7\times 10^{10}$ OP\,s$^{-1}$W$^{-1}$ (operations per second per watt). This estimation does not take into account the energy cost of external memory access, which is crucial in neural networks simulations and can be much higher than the cost of the computation itself. Memory access is an important bottleneck in CMOS neural network implementations~\cite{Xu_EdgeInference} with the minimum cost per bit at the level of 10\,pJ. The efficiency of $10^{13}$\,OP\,s$^{-1}$W$^{-1}$ was reported for the state-of-the-art energy-optimized neural network accelerator~\cite{Lee_UNPU}, but such a high efficiency was reported for convolutional layers of the network only, which are characterized by a relatively small number of external memory access operations. In contrast, in our neuromorphic design the cost of memory access is zero, as there is no need to store the state of the system in external memory during computation. Finally, we mention that off-the-shelf GPU systems~\cite{Canziani} can reach the energy efficiency at the level of $10^9$ OP\,s$^{-1}$W$^{-1}$.

In some other works~\cite{Chen_ClassificationSilicon,Zhou_LargeScale}, efficiency of hardware neural networks was determined by estimating the number of FLOPs required to simulate the network rather than comparing the number of SOPs. Such an approach can give estimates orders of magnitude higher but here we choose to take a more conservative approach by focusing on the number of achievable SOPs. In the case of a disordered dopant atom network~\cite{Chen_ClassificationSilicon}, it was suggested that a theoretical efficiency limit of $10^{13}$\,SOP\,s$^{-1}$W$^{-1}$ could be achieved~\footnote{In~\cite{Chen_ClassificationSilicon} 1 SOP = 10 FLOPs conversion ratio was assumed.}, but this estimation does not account for the cost of the output signal thresholding and additional processing of the output, which have to be performed electronically in this case. The comparison with other neuromorphic and non-neuromorphic systems is presented in Fig.~\ref{fig:comparison}, which also indicates the efficiency of our recent experimental implementation of a network based on a binarized polariton node with non-resonant pumping~\cite{Mirek_Neuromorphic}. According to this comparison, we estimate that an exciton-polariton network has the potential to outperform other solutions by orders of magnitude. In the above comparison we do not include estimations made for optical systems that perform linear operations (in the function of inputs), such as vector-matrix multiplication~\cite{Soljacic_DeepLearning,Xu_TOPS,Feldmann_parallel,Chang_CNN,nahmias2019photonic,Lin,Zhou_LargeScale}, since such systems alone are not able to efficiently perform complex machine learning tasks that require nonlinearity. 

\subsection{Footprint and performance density} \label{sec:footprint}

To estimate footprint, we consider the minimal physical size of optical weight banks which encode synaptic weights. The size of microcavity neurons is comparable to the size of weight banks, but their number is smaller. The size is limited by the optical wavelength and typically cannot be decreased below micrometer level. We cautiously assume the size of 10 $\mu$m$^2$ for both the optical weight and optical  node. Each weight can be realized with a semi-transparent filter, either embedded within a spatial light modulator array, or even with a non-adjustable filter such as a patterned glass surface, if the weights can be set at the fabrication stage.

While this size is much larger than the size of an electronic transistor, it is unlikely to be the main limiting factor for the footprint. One million polariton neurons or optical  weights can be implemented on a 10 mm$^2$ area, which is comparable to the density in a neuromorphic CMOS architecture~\cite{Merolla}. The number of parameters in the leading artificial neural networks participating in the ImageNet competition~\cite{Xu_EdgeInference} is of the order of $10^8$. Such number of parameters would require an optical weight bank surface of a few cm$^2$.

The most important performance measure is the performance density, which is the number of operations that can be performed per surface area per time~\cite{Xu_EdgeInference}. The practical maximum rate of incident pulses in a polariton system system can be estimated to be of the order 1/(100 ps). After injecting the input pulses, a typical polariton system needs the ``cooling-off period'' of the order of 100 ps to completely recover to the original state, so that the response of the cavity is not affected by the previous pulses~\cite{Mirek_Neuromorphic}. This rate is given by the lifetime of reservoir excitons, or unwanted excitations in the quantum well, and may strongly depend on the material used. Such excitations occur even in the case of resonant excitation. Here, we assume that the cavity polariton lifetime is shorter than 100 ps. The resulting performance density is orders of magnitude higher than in the leading CMOS systems~\cite{Xu_EdgeInference}. Figure~\ref{fig:comparison} presents comparison of performance density with respect to other systems. 

However, the performance density does not necessarily tell the full story. It is also important to consider the aspect of heat generation in the calculation of footprint. In electronic devices, due to  heat generation, it is difficult to implement multi-layer structure in three-dimensional chips. One of the dimensions is sacrificed for a heat drain, and it is the reason why performance density is measured in operations per surface and not per volume. Hence, calculation of footprint should not be considered without the consideration of energy dissipation, which is also here the ultimate limiting factor. In the case of a system with much lower energy loss but larger surface area, it is possible to consider multi-layer wiring, exploiting the third dimension, which is typically occupied by a heat sink. For example, a physical multi-layer network can be considered, where each layer is placed on top of the previous one. In the following section we discuss the possible integration of the proposed system inside an optical chip.

\subsection{Extensions and generalizations} \label{sec:other}

\begin{figure}
    \includegraphics[width=\linewidth]{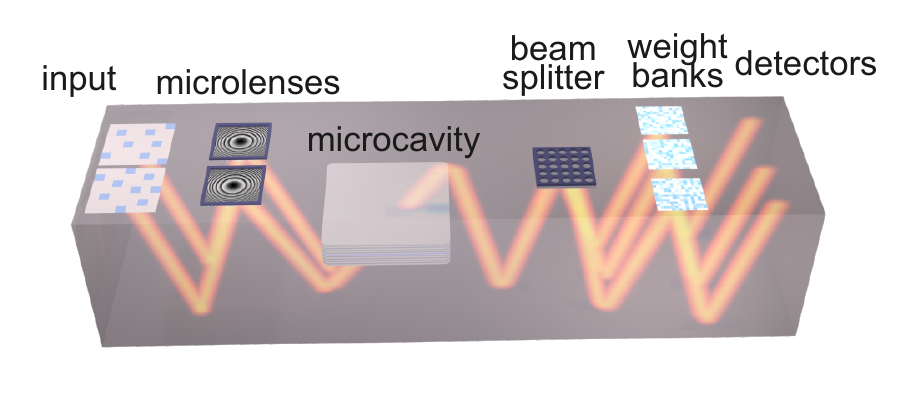}
    \caption{Scheme of a possible integrated version of the free-space system. The subsequent layers of the system are arranged side by side in the plane of a planar waveguide. Elements such as lenses and beam splitters can be implemented with diffractive optical elements.}
    \label{fig:integrated}
\end{figure}

We consider several possible extensions of the proposed design. The system depicted in Fig.~\ref{fig:design} is based on free-space optics, but an integrated version is required for a portable device. We note that the idea behind free-space optics does not necessarily imply propagation in free space. The same principles can be applied to design an integrated system, where light propagates in an optical medium, such as glass or a transparent semiconductor. In fact, the most common integrated devices that rely on free-space optics are mobile phone cameras. Millimeter-sized camera objectives are used together with integrated semiconductor arrays, where each optical sensor is micrometer-sized. Analogously, it is possible to design portable neural networks based on micrometer-sized nodes and focal lengths of the order of millimeters. Figure~\ref{fig:integrated} illustrates the possible geometry of an integrated polariton neural network. A device relying on a similar concept has been demonstrated in Ref.~\cite{gruber2000planar}. Propagation is in this case along the cavity plane and lenses are replaced by diffractive optical elements, which can be designed to precisely control light paths. 

An alternative to free-space propagation is the use of integrated optical waveguides~\cite{Feldmann_AllOpticalSpikingNetwork,Soljacic_DeepLearning}. Waveguide systems can be implemented in a compatible semiconductor platform, and can be used not only to transport light, but also to build efficient devices for linear operations, such as convolutions or vector-matrix multiplication~\cite{Soljacic_DeepLearning,Feldmann_parallel}. It is also possible to consider a hybrid architecture where light propagation is partly based on free-space optics and partly on light confinement in waveguides. Waveguide systems have two potential drawbacks: they suffer from photon losses, in particular at waveguide bends, and require a separate waveguide for each optical path. On the other hand, free-space connectivity can be almost loss-free, and allows arbitrary intersection of optical paths in a linear medium, which can greatly improve data bandwidth~\cite{Miller_Attojoule,Miller_device}.

Another interesting possible extension is the use of holography for encoding synaptic weights~\cite{Psaltis_Holography}. In this case, data can be efficiently stored in a three-dimensional, rather than two-dimensional, weight bank. This can improve memory capacity significantly, but on the other hand, writing and retrieving information requires complicated optical systems.

A straightforward generalization of our design is to use neurons with analog, rather than binary, inputs and outputs.  In this work we chose the simple XOR gate as an example of a neuron which already allows to perform complex machine learning tasks. From the point of view of machine learning, the XOR gate, in contrast to OR, AND gates, is itself a classification task that requires nonlinearity to be solved~\cite{Mirek_Neuromorphic}. Moreover, it is a sufficiently strongly nonlinear function that it allows to be used as a basic building block for the construction of more complex networks~\cite{Rastegari,Bengio_Binarized,Mirek_Neuromorphic}. In the case of analog neurons, a similar condition exists. The activation function has to be strongly nonlinear, with a negative differential response~\cite{Mirek_Neuromorphic}, in order to be useful for efficient solution of complex machine learning problems. Optical weights may also modify phases of optical pulses rather than intensities, which can reduce optical losses and increase the available parameter space. 

An extension to a multi-layer system is possible, thanks to the resonant character of the input, which means that input and output pulses have the same optical frequencies. In this case, multiple weight banks apply the synaptic connections between neurons in subsequent layers, and each neuron layer can be implemented with a separate microcavity. On the other hand, in this case losses become a more serious issue and may accumulate exponentially as light is transmitted through a number of layers. In Sec.~\ref{sec:supertransparent} we show how an appropriate design of a microcavity can circumvent losses to a large extent. Nevertheless, in the case of a large number of layers, some kind of pulse regeneration method will likely be required. Note that difficult tasks such as ImageNet can be solved very efficiently with networks containing only a few layers, for example by AlexNet (8 layers). 
 
Finally, the vector-matrix multiplication can be replaced by a convolution operation~\cite{Feldmann_parallel,Chang_CNN}. This is particularly important for image recognition tasks, where convolutional layers play a very important role~\cite{LeCun_MNIST}, or in general in tasks that are characterized by translational invariance, either in space or in time.

\section{Implementation of binarized neurons} \label{sec:methods}

\begin{figure}
    \includegraphics[width=\linewidth]{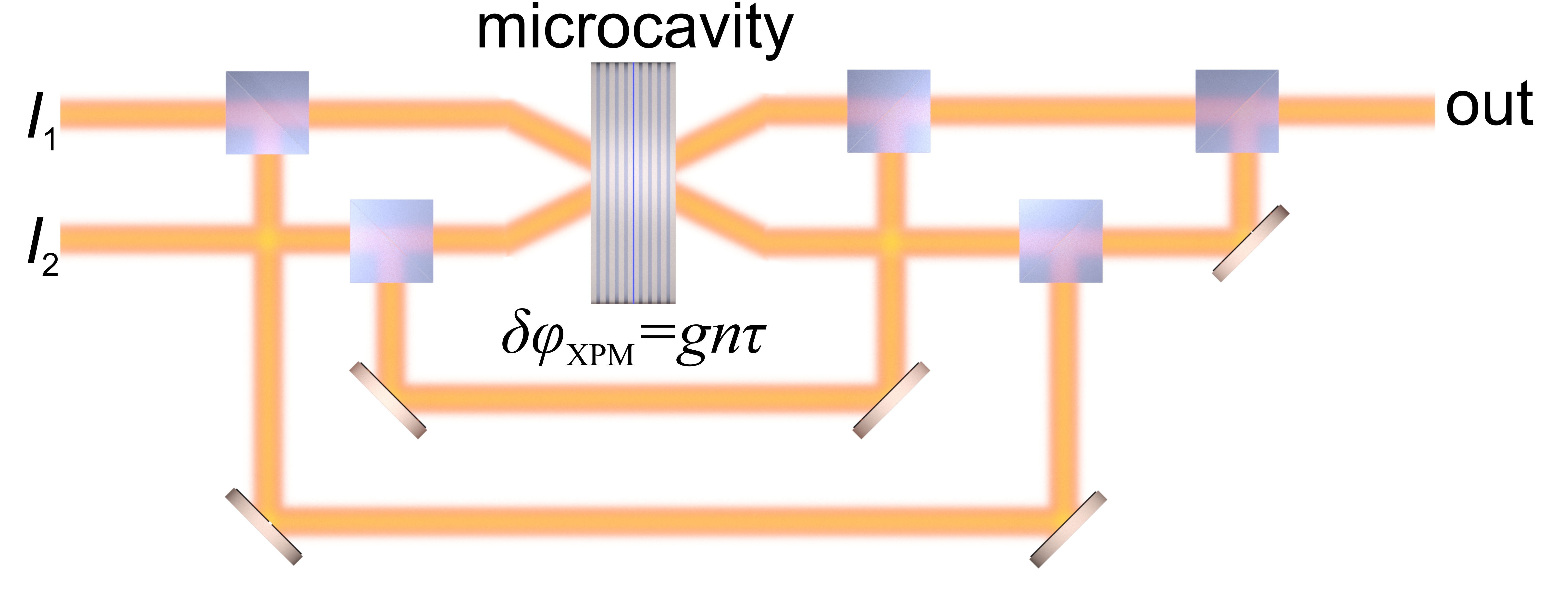}
    \caption{Scheme of a possible XOR gate implementation. The principle of operation is based on cross phase modulation (XPM) of two pulses inside the microcavity and interferometry. With appropriate tuning of optical path lengths, interference of cavity emission with copies of input pulses results in intensity minimum for the "11" input configuration.}
    \label{fig:gate_MZI}
\end{figure}

We propose two methods of implementation of XOR gates in a microcavity system in the case when input pulses are quasi-resonant with the frequency of the microcavity. Resonant excitation is better suited to energy efficient processing than the nonresonant excitation method used in the recent experiment~\cite{Mirek_Neuromorphic}, as in the latter case input pulses do not create exciton-polaritons directly. Instead, high-energy excitations in the so-called reservoir are created, and only a fraction of them is converted into polaritons. Moreover, with the resonant excitation method, the frequency of the input and the output are equal, which facilitates realization of multi-layer networks. However, spectral filtering used in~\cite{Mirek_Neuromorphic} cannot be directly used to obtain a negative response in the case of resonant excitation, as the frequency of photons is fixed.

The first method is based on cross-phase modulation and the use of interferometers. We assume that ``0'' is encoded with the absence of a pulse or a low-intensity pulse, and ``1'' with a high-intensity pulse. Before entering the microcavity, each of the input pulses is split into two equal copies, see Fig.~\ref{fig:gate_MZI}. One of the two copies is sent around the microcavity, while the other copy is incident on the surface of the microcavity. The second input pulse is split and directed in the same way. The copies of the two input pulses meet at the microcavity, where they interact through polariton-polariton scattering. This induces both self-phase modulation, which is independent of the other pulse, and cross-phase modulation, which is dependent on the intensity of the other pulse. The phase acquired by cross phase modulation is proportional to the interaction energy between polaritons in the two pulses, i.e.~each pulse $i=1,2$ acquires the phase $\delta\phi_i\approx g_{\rm XPM}\langle n_{3-i}\rangle\tau$, where $\langle n_{3-i}\rangle$ is the average density of polaritons in the other pulse during the time of interaction, which is given by $\tau$. As the two pulses are incident at different angles, they separate at the exit from the microcavity. Each of the transmitted pulses is then combined with its copy that did not go through the microcavity, which results in interference. Interference is destructive when the cross-phase modulation phase $\delta\phi$ together with the phase difference accumulated along optical paths is equal to $\pi$. By adjusting the delay lines between the interfering pulses accordingly, it is possible to obtain destructive interference for input pulses that are in the "11" state.

As we are only interested in order of magnitude estimations, we neglect the temporal dynamics of $n$ and assume that it is an average of density calculated over the duration of the interaction $\tau$. For optimal interaction, the temporal length of the pulse and the lifetime of polaritons in the microcavity should be similar. This ensures that polaritons can be effectively created in the microcavity and most of them can be present and interact there at the same time. It also ensures that the spectral density of the pulse matches the width of the spectral resonance of the microcavity. This requirement leads to the condition $gn = \hbar \gamma$ mentioned in Sec.~\ref{sec:estimates}.

\begin{figure}
    \includegraphics[width=\linewidth]{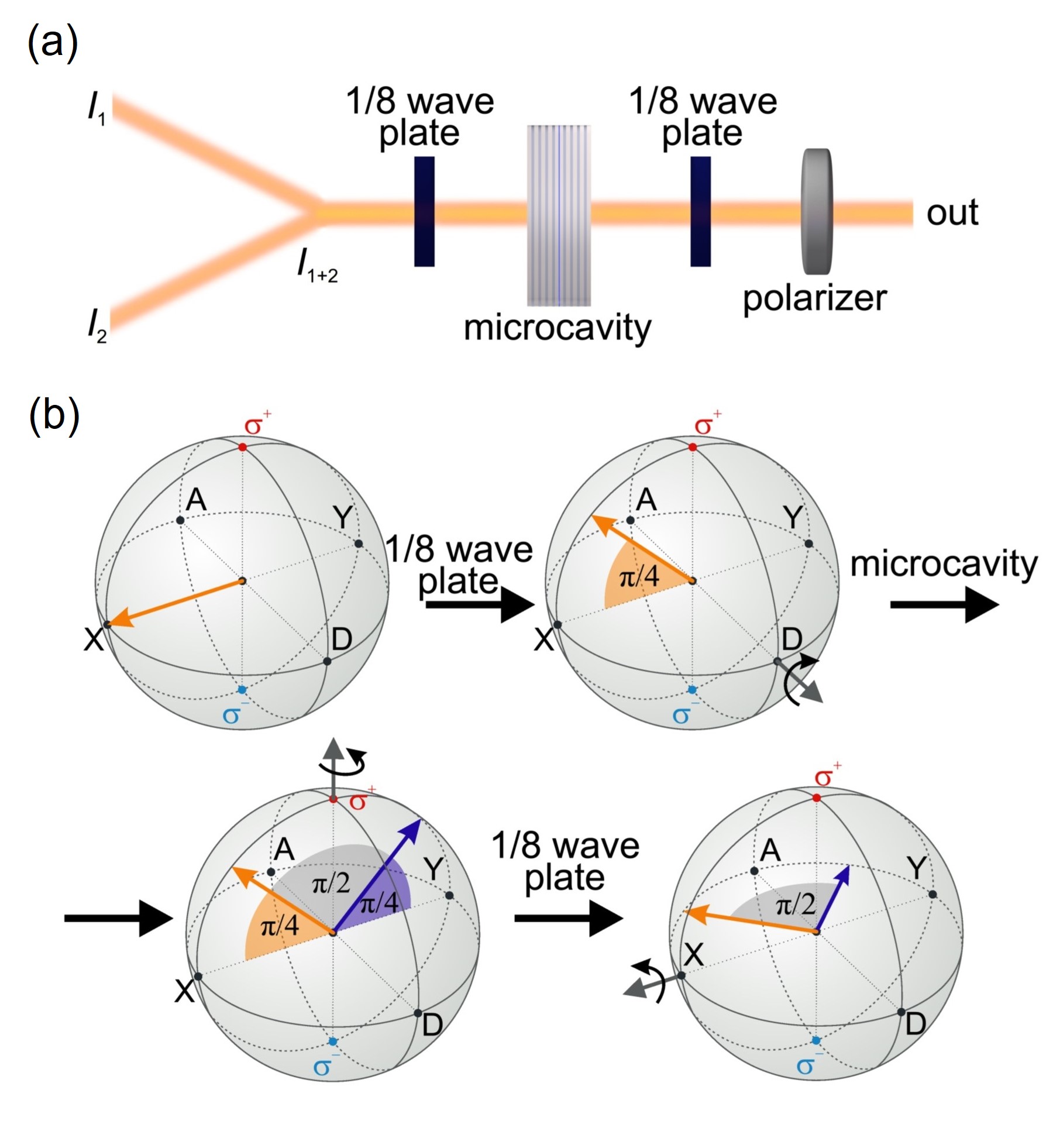}
    \caption{(a) Scheme of the XOR gate setup based on polarization rotations. (b) Details of rotations visualized on the Poincar\'e sphere. The input pulses are assumed to be linearly polarized. The first \sfrac{1}{8}-wave plate rotates the polarization by the angle $\pi/4$ around the diagonal-antidiagonal polarization axis. The spin-anisotropic polariton interaction results in the rotation of polarization around the circular polarizations axis, which is intensity dependent. This results in $\pi/2$ angle between polarizations of the outputs from the microcavity in the input configurations "10/01" (orange) and "00/11" (purple). The second \sfrac{1}{8}-wave plate rotates the polarization back to the equator (linear polarization) plane. The polarizer at the output blocks the polarized emission corresponding to the "00/11" configuration.}
    \label{fig:gate_spinor}
\end{figure} 

The drawback of the above method is that it uses an interferometric technique, which requires coherence of light and precise adjustment of optical path lengths for all nodes, which may be impractical. These shortcomings are not present in the second method, which utilizes the anisotropy of interactions of exciton-polaritons~\cite{Kavokin_Book}. In this case, we assume that the two input pulses are mutually incoherent. Interactions between polaritons with opposite spin (or circular polarization) are very weak, while polaritons with the same spin interact strongly~\cite{Vladimirova_InteractionConstants}. This allows to construct a simple XOR gate using two \sfrac{1}{8}-wave plates, one placed in front and one behind the cavity, and a polarizing filter. The principle of the method is shown in Fig.~\ref{fig:gate_spinor}. In this case, the two input pulses are assumed to be linearly polarized in a well defined direction (X). At the entrance, they are combined into one. Before entering the microcavity, the combined pulse goes through a \sfrac{1}{8}-wave plate, which rotates the polarization by 45 degrees on the Poincare sphere. Inside the microcavity, the polaritons interact, which leads to nonlinear spin rotation around the $z$ axis on the Poincare sphere, which is dependent on the intensity of the input pulse. We assume that the intensity of the pulse in the "10" or "01" input combination leads to a rotation that is different by $\pi$ with respect to the "00/11" combination. As in Sec.~\ref{sec:estimates}, this requires that the density and time of interaction fulfill the condition $g n \tau \approx 1$. After exiting the cavity, the polarization is rotated again by 45 degrees using a \sfrac{1}{8}-wave plate back to the equator plane, which results in linear polarization of pulses, orthogonal in the case of "01/10" and "00/11" polarizations. The pulse now passes through the polarizer, which blocks the light corresponding to the "00/11" polarization, which results in an XOR gate output. A similar concept was proposed to realize a spin transistor~\cite{Shelykh_SpinTransistor}, where spin rotation due to the polariton energy splitting followed by interference was used.

\section{Microcavity transparency} \label{sec:supertransparent}

\begin{figure}
    \centering
    \includegraphics[width=\linewidth]{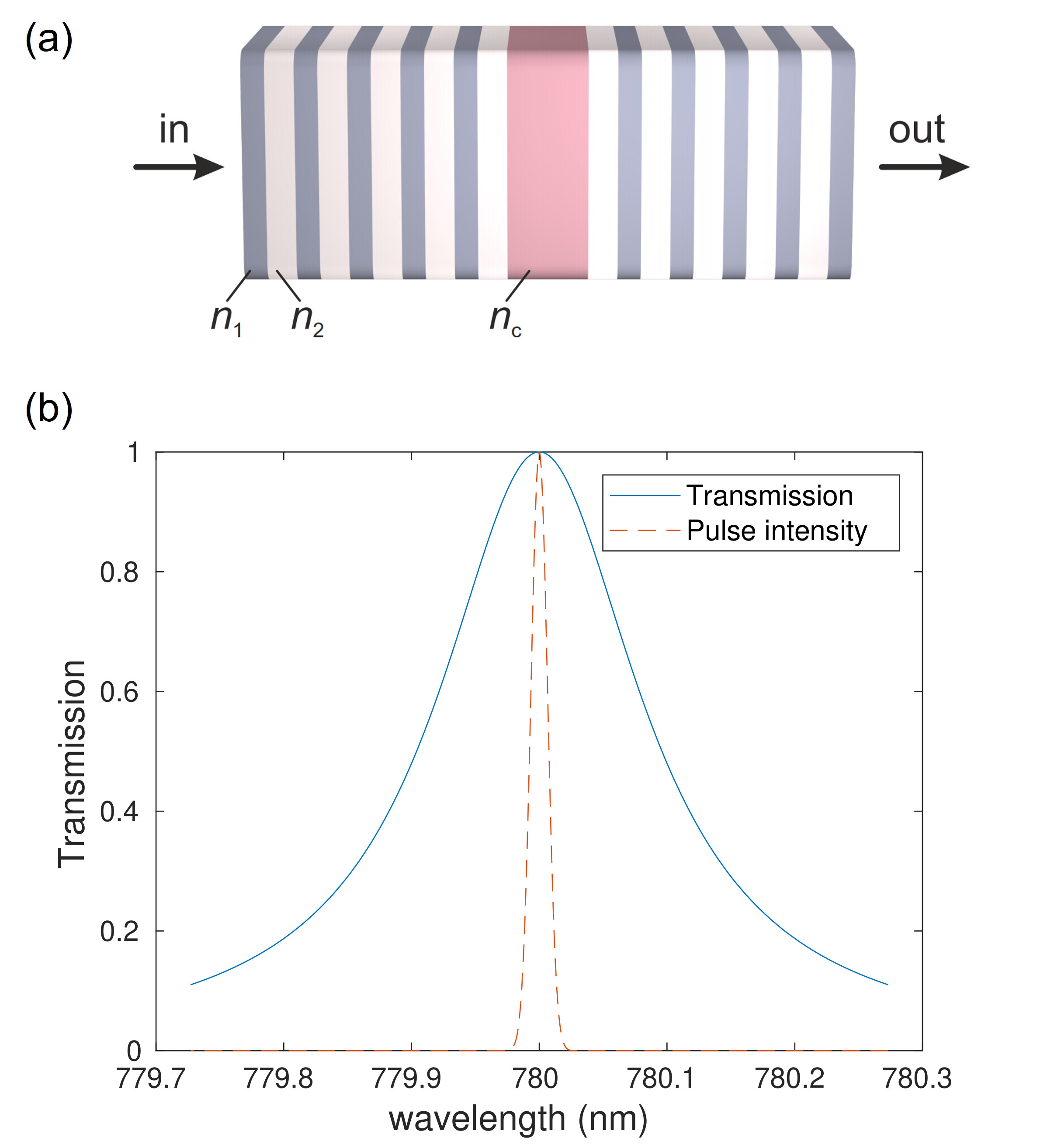}
    \caption{(a) Example of a microcavity considered in the text. (b) Solid line shows numerically calculated transmission spectrum of a microcavity with mirrors composed of 25 pairs of layers each with refractive indices $n_1=3$ and $n_2=3.5$ and the cavity with refractive index $n_{\rm c}=3$. Widths of the mirror layers are $\lambda/4n_i$, where $\lambda$ is the central wavelength, and the width of the cavity is $\lambda/2n_{\rm c}$. Dashed line shows the normalized spectrum of a laser pulse with 50 ps temporal FWHM.}
    \label{fig:cavity}
\end{figure}

Efficient operation of the proposed neural network requires low energy losses at optical elements. On the other hand, to enhance polariton-polariton interactions, microcavities with very high quality factor are required. The use of high-reflectivity mirrors in such cavities may seem to contradict the assumed high transmission coefficients as in Table~\ref{table}. However, this is not the case, and in this Section we show that high-Q cavities can be characterized by high transmission rates at resonance.

Consider transmission of a light pulse through a dielectric microcavity structure depicted in Fig.~\ref{fig:cavity}, where mirrors consist of alternating layers of materials with refractive indices $n_1$ and $n_2$, and the cavity medium has the index $n_{\rm c}$. The coefficients of transmission and reflection are conveniently calculated using the transfer matrix formalism~\cite{Kavokin_Book}. 
At the center of the stop band we assume $n_j d_j=\lambda/4$, where $\lambda$ is the light wavelength in vacuum, $d_j$ is the thickness of the mirror layer $j$, and the transfer matrix is
\beq
{\bf T}_j = 
\begin{pmatrix}
  0 &  i/n_j  \\
  i n_j & 0
\end{pmatrix}.
\eeq
Two layers of material with refractive indices $n_1$ and $n_2$ stacked next to each other correspond to the transfer matrix which is the product of ${\bf T}_2$ and ${\bf T}_1$ 
\beq
{\bf T}_{12} = 
\begin{pmatrix}
  -n_1/n_2 & 0 \\
  0 & -n_2/n_1
\end{pmatrix}.
\eeq
Perfect transmission can be achieved when the complete transfer matrix of the microcavity is proportional to the identity matrix, cf.~Fig.~\ref{fig:cavity}
\beq
    {\bf T}_{\rm mc} = {\bf T}_{12}^N {\bf T}_{\rm c} {\bf T}_{21}^N \sim
    \begin{pmatrix}
  1 &  0  \\
  0 & 1
\end{pmatrix},
\eeq
where $N$ is the number of dielectric layer pairs on each side. The above condition is fulfilled when the width of the cavity layer is such that $n_{\rm c} d_{\rm c}=\lambda/2$.

Bottom panel of Fig.~\ref{fig:cavity} shows the transmission spectrum of a microcavity composed of 25 pairs of layers on each side of the microcavity, and fulfills the above condition. Despite the high Q-factor of the cavity and high reflectivity of the mirrors, almost perfect transmission occurs at the resonant frequency. Similarly, a polariton cavity containing quantum well excitons can be designed to exhibit such a high transmission at resonance.

In practice, effects such as cavity imperfections, absorption, and nonradiative exciton decay will decrease the peak transmission rate. Currently, polariton cavities can reach transmission coefficients at the level of a few tens of percent.

\section{Neural networks versus universal computing}

We would like to point out several advantages of our neural network design with respect to the more conventional universal digital logic. 
Miller~\cite{Miller_OpticalLogic} summarized the difficulties in using optical components for information processing and pointed out conditions that have to be fulfilled by a reasonable candidate for digital logic. However, some of these conditions, including logic level restoration condition, or ``cleaning up'' the signal, and the fan-out condition are only necessary in the implementations that follow the architecture of conventional computers. In particular, they are not necessary in the case of single hidden layer networks implemented in hardware. They are also not absolutely necessary for deep feed-forward networks, if optical signals are not redirected recurrently.

Nevertheless, in our system some of the conditions of~\cite{Miller_OpticalLogic} are fulfilled: (a) critical biasing is not an issue since the input-output operating point is not threshold-like;  (b) input-output isolation can be achieved by a simple method of directing the input pulses in free space at an angle with respect to optical surfaces, which results in reflected pulses propagating along different trajectories than the input pulses; (c) the condition of cascadability can be fulfilled in the case of resonant excitation, when the frequency of the output is the same as the input, provided that some method of optical signal restoration or amplification is implemented~\cite{Bloch_1DAmplification}; (d) while the logic level is dependent on loss in our system, losses can be minimized as described in Sec.~\ref{sec:supertransparent}, and are proportional in each layer-to-layer connection, which enables to define signal amplitudes for logic levels accordingly in each layer. At the same time, it is possible to correct for the difference in loss in different optical paths by adjusting the corresponding optical weights. 

The use of a feed-forward neural network architecture instead of universal digital computing results in another important aspect of the system, which is the absence of separate memory units. Currently, there is no reliable implementation of optical memory that could be used to implement a traditional computing architecture with polaritons. Even if such technology existed, it would be required to store and read information with ultrashort, ultra-low energy pulses to comply with the energy efficiency of the entire system, which may be difficult to realize in practice.

\section{Quantum limit of energy efficiency} \label{sec:quantum}

Since our device is based on classical physics, it should operate well above the quantum limit to avoid quantum noise. It is interesting to compare the potential energy efficiency of our system to the single-photon limit, since the fundamental energy efficiency bound is imposed, as in the case of electronics, by quantum effects. The quantum limit corresponds to a single photon per bit of information or two photons per XOR gate. This leads to the estimation of 10$^{18}$\,SOP\,s$^{-1}$W$^{-1}$, as depicted in Fig.~\ref{fig:comparison}. The onset of the quantum limit has been already observed in tightly confined exciton-polariton microcavities~\cite{Volz_QuantumCorrelatedPhotons,Imamoglu_TowardsPolaritonBlockade}. We would like to emphasize that even if XOR nodes operated close to the quantum limit, few-photon emitters or detectors would not be needed. The ``quantum limit'' in the sense considered here applies to the operation of network nodes in the hidden layer. Since there are no separate detectors or emitters in these nodes, we don't need to be concerned with the detection of these weak signals. As already noted, output pulses arriving at the detectors in the final layer can have intensities orders of magnitude higher than the pulses that perform individual operations in the hidden layer nodes.

\section{Discussion}

In this work we analyzed the potential energy efficiency of all-optical exciton-polariton based systems for information processing and proposed a design of a simple neural network for data classification. The advantage of performance density and energy efficiency is achieved in the inference, while the training stage is assumed to be performed in software. The software model of a neural network resulting from training can be implemented in optical hardware by setting the appropriate optical weights. In this sense, the proposed system is an example of an optical neural network accelerator~\cite{Chen_DianNao,Lee_UNPU,Kitayama_PhotonicAccelrator}, where previously trained software model is implemented in hardware to increase the performance of inference. Such systems are best suited to machine learning tasks where analysis of a large number of samples is required, while re-training or replacement of the model is not necessary or performed unfrequently. Such tasks include image, speech and video recognition, natural language processing, automatic detection and control systems.

A practical system is required to operate at room temperature. While the majority of laboratory experiments with exciton-polaritons are performed at cryogenic environments, there are already many experiments with room-temperature exciton-polariton systems, including inorganic~\cite{Grandjean_RTPolaritonLasing,Malpuech_ZnOCondensate} and organic~\cite{Plumhof_NMat_2014, Daskalakis_NMat_2014, Dusel_NComm_2020} semiconductors, two-dimensional materials~\cite{Tartakovskii_vdWPolaritons} and perovskites~\cite{Fieramosca_perovskites,Su_Perovskites}. These materials are typically characterized by a more prominent sample disorder, but if scattering on disorder is not strong enough to make the signal in the output too weak, it can be part of the internal losses that do not alter the efficiency of the operation. On the other hand, the variation of node response due to disorder and variability from one sample to another can be corrected for by appropriately adjusting optical weights at the input and output layers. At the same time, we expect that heating will be a negligible effect due to the low optical powers required. Effects of heating are not typically observable in experiments with polaritons, except for the highest pumping rates.

The possible sources of optical losses are expected to be associated with imperfection of cavity mirrors and irreversible (polariton unrelated) absorption of photons in the sample. The technology of microcavity fabrication is sufficiently advanced to produce very clean mirrors, which results in extremely high optical mode quality factors. Irreversible absorption may result in sample heating, however this effect is expected to be negligible at low optical pulse energies. The creation of a long-lived exciton reservoir by resonant laser pulses is also possible~\cite{Walker_DarkSolitons}, but this effect becomes less important for ultrashort pulses. Nevertheless, we accounted for the exciton reservoir lifetime in our estimation of maximum possible data rate in Sec.~\ref{sec:footprint}. The use of picosecond pulse laser sources  is required, which contributes to the overall size of the system. The possible alternative to bulky ultrashort pulse lasers is the use of compact VCSEL lasers together with ultrafast optical modulators to shape picosecond input pulses~\cite{Hurtado_Ultrafast}.

\acknowledgments

MM acknowledges support from National Science Center, Poland grant 2017/25/Z/ST3/03032 under the QuantERA program. AO acknowledges support from National Science Center, Poland grant 2016/22/E/ST3/00045. RM acknowledges support from National Science Center, Poland grant 2019/33/N/ST3/02019. BP acknowldges support from National Science Center, Poland grant 2020/37/B/ST3/01657. KT acknowldges support from National Science Center, Poland grant 2020/04/X/ST7/01379.  TL acknowledges the support of the Singapore Ministry of Education, via the Academic research fund project MOE2019-T2-1-004.

\bibliography{bibliography}

\end{document}